\documentclass[11pt]{article}
\usepackage{amssymb,latexsym,amsmath}
\usepackage[dvips]{graphicx}
\newtheorem{theo}{Theorem}

\newtheorem{prop}[theo]{Proposition}
\def\ds{\displaystyle}
\def\nn{\nonumber}
\def\deg{\mathop{\rm deg}\nolimits}
\def\str{\mathop{\rm str}\nolimits}
\def\qdots{\mathinner{\mkern1mu\raise1pt\vbox{\kern7pt\hbox{.}}\mkern2mu
 \raise4pt\hbox{.}\mkern2mu\raise7pt\hbox{.}\mkern1mu}}
\def\Z{{\mathbb Z}}

\def\gl{\mathfrak{gl}}
\def\ssl{\mathfrak{sl}}
\def\h{\mathfrak{h}}
\def\so{\mathfrak{so}}
\def\osp{\mathfrak{osp}}
\def\a{\alpha}
\def\b{\beta}
\def\J{{\bf J}}
\def\hR{\hat{R}}
\def\hP{\hat{P}}
\def\hJ{\hat{J}}
\def\hbR{\hat{\bf R}}
\def\hbP{\hat{\bf P}}
\def\lb{[\![}
\def\rb{]\!]}

\begin{document}
\begin{center}
{\Large \bf
Representations of the Lie Superalgebra $\gl(1|n)$ in a \\[2mm]
Gel'fand-Zetlin Basis and Wigner Quantum Oscillators}\\[5mm]
{\bf R.C.~King}\footnote{E-mail: R.C.King@soton.ac.uk}\\[1mm]
School of Mathematics, University of Southampton,\\
Southampton SO17 1BJ, U.K.;\\[2mm]
{\bf N.I.~Stoilova}~\footnote{E-mail: Neli.Stoilova@UGent.be; Permanent address:
Institute for Nuclear Research and Nuclear Energy, Boul.\ Tsarigradsko Chaussee 72,
1784 Sofia, Bulgaria} {\bf and J.\ Van der Jeugt}\footnote{E-mail:
Joris.VanderJeugt@UGent.be}\\[1mm]
Department of Applied Mathematics and Computer Science,
Ghent University,\\
Krijgslaan 281-S9, B-9000 Gent, Belgium.
\end{center}

\vskip 10mm


\begin{abstract}
An explicit construction of all finite-dimensional irreducible representations of 
the Lie superalgebra $\gl(1|n)$ in a Gel'fand-Zetlin basis is given. 
Particular attention is paid to the so-called star type~I representations (``unitary representations''),
and to a simple class of representations $V(p)$, with $p$ any positive integer.
Then, the notion of Wigner Quantum Oscillators (WQOs) is recalled.
In these quantum oscillator models, the unitary representations of $\gl(1|DN)$ are 
physical state spaces of the $N$-particle $D$-dimensional oscillator. 
So far, physical properties of $\gl(1|DN)$ WQOs were described only in the so-called Fock
spaces $W(p)$, leading to interesting concepts such as non-commutative coordinates and
a discrete spatial structure. 
Here, we describe physical properties of WQOs for other unitary representations,
including certain representations $V(p)$ of $gl(1|DN)$.
These new solutions again have  remarkable properties following from the spectrum
of the Hamiltonian and of the position, momentum, and angular momentum operators.
Formulae are obtained that give the angular
momentum content of all the representations V(p)
of $\gl(1|3N)$, associated with the $N$-particle 3-dimensional WQO.
For these representations $V(p)$ we also consider in more detail the
spectrum of the position operators and their squares, leading to interesting consequences.
In particular, a classical limit of these solutions is obtained, that is in agreement
with the correspondence principle.
\end{abstract}

\vfill\eject

%

\setcounter{equation}{0}
\section{Introduction} \label{sec:Introduction}%

In this paper we construct all finite-dimensional irreducible representations 
of the general linear Lie superalgebra $\gl(1|n)$~\cite{Kac1,Kac2}. 
To this end we introduce a Gel'fand-Zetlin basis (GZ-basis)
for $\gl(1|n)$ and write down explicit expressions for the transformation of 
the basis vectors under the action of the algebra generators. 
This is analagous to the GZ-basis for $\gl(n|1)$ and the action of its generators 
given by Palev~\cite{Palev2}.
The Lie superalgebra $\gl(1|n)$ is a central extension of the special linear 
Lie superalgebra $\ssl(1|n)$. Each finite-dimensional irreducible 
$\gl(1|n)$ representation remains irreducible when restricted to $\ssl(1|n)$, and
each of the resulting finite-dimensional irreducible representations of $\ssl(1|n)$ 
is either typical or atypical~\cite{Kac1,Kac2}. 

The motivation for the present work stems from the fact that a set of generating 
elements of $\ssl(1|n)$ satisfies the compatibility conditions of the so called 
Wigner Quantum Oscillator (WQO)~\cite{Wigner}-\cite{SJ}. 
Therefore to investigate the properties of these 
$\ssl(1|n)$ WQOs we must consider those representations that are of physical relevance, 
namely the star representations of type I~\cite{Gould}. These include certain 
representations $W(p)$~\cite{Palev3} having highest weight $\Lambda=(p;0,0,\ldots,0)$,
that have a Fock space realisation. 
A study of these in the case $n=3$ and $n=3N$ has revealed
WQO models that have a finite, equally spaced energy spectrum, with discrete
values of measurements of both spatial coordinates and linear momenta, as well
as angular momenta~\cite{Palev1, Palev5, K1,K2}. Moreover, the underlying geometry of the model is non-commutative~\cite{HS, C},
in the sense that coordinate operator components in general do not commute;
thus measurements of two different coordinates, say $x$ and $y$, 
of a single particle cannot be performed simultaneously. 
As such, the WQO models are to be
compared with other non-commutative models, see for example~\cite{NP}-\cite{MM}.

Here the intention is to set up the mathematical machinery enabling these 
properties to be explored for WQO models based on any $\ssl(1|n)$ star representation of type~I.
The results confirm that in all such cases there is a finite number of equally
spaced energy levels, that spatial coordinates and linear momenta are
discretely quantised, and that the model always exhibits non-commutative
geometry. A detailed analysis is presented for those WQO models based on
the $\ssl(1|n)$ representations $V(p)$ having highest weight $\Lambda=(1;p-1,0,\ldots,0)$,
and also for a representation of highest weight $\Lambda=(2;1,0,\ldots0)$
that belongs to neither class $W(p)$ nor $V(p)$.

The structure of the paper is as follows. In Section~2 we construct all the 
finite-dimensional irreducible representations of the Lie superalgebra $\gl(1|n)$, complete
with a specification of the GZ basis vectors $|m)$ and the explicit action of a set of 
$\gl(1|n)$ generators on these vectors. The section culminates with formulae
for the action on an arbitrary vector $|m)$ of all the odd elements, $e_{0j}$ 
and $e_{j0}$, of $\gl(1|n)$. These are of importance in Section~4, where 
for $n=DN$ these odd elements play the role of certain linear 
combinations of position  and linear 
momentum operators for the $N$-particle $D$-dimensional WQO. 
Before this, in Section~3, the formulae of Section~2 are applied to the case of 
the $\gl(1|n)$ representations $V(p)$ of highest weight $\Lambda=(1;p-1,0,\ldots,0)$.
The GZ-basis and the action of the $\gl(1|n)$ generators are given explicitly in 
a rather more succinct notation than was possible in the general case.

Returning to Section~4, it is here that the WQO is introduced. The Hamiltonian
$\hat{H}$ of an $N$-particle $D$-dimensional harmonic oscillator takes the form
\begin{equation}
\hat{H}=\sum_{\alpha=1}^{N} \Big( \frac{\hbP_\alpha^2}{2m}
+ \frac{m\omega^2}{2} {\hbR}_\a^2 \Big), \label{Intro-H}
\end{equation}
with $\hbP_\alpha$ and $\hbR_\alpha$ $D$-dimensional vector operators
corresponding to the momentum and position of the particle~$\alpha$, each
of the same mass $m$ and natural frequency $\omega$. This 
Hamiltonian can be re-expressed in the form
\begin{equation}
     \hat{H}= \frac{\hbar\omega}{DN-1} \sum_{j=1}^{DN} \{A_j^+, A_j^-\},
     \label{Intro-HAA}
\end{equation}
where the operators $A_j^\pm$ are simple linear
combinations
of $\hat{R}_{\a k}$ and $\hat{P}_{\a k}$ for some $\a$ and $k$ determined by
$j$.
With this notation, 
the WQO requirement that Hamilton's equations and the Heisenberg equations 
coincide as operator equations leads to compatibility conditions
on the operators $A_j^+$ and $A_j^-$ that have a non-canonical solution
allowing them to be identifed, as stated above, with the odd generators,
$e_{j0}$ and $e_{0j}$, respectively, of $\gl(1|n)$. This identification
is then exploited to determine the physical properties of WQO models,
including their energy spectrum and the eigenvalues of their spatial 
coordinate operators, as well as their non-vanishing
commutator. The determination of the angular momentum content 
of the particular $\gl(1|3N)$ modules $V(p)$ is deferred to Section~5, 
where it is achieved for all $N$ and all $p$, by means of a generating 
function derived from Molien's Theorem~\cite{Molien, Sturmfels} applied to the case
of the embedding of the rotation group $SO(3)$, with Lie algebra $\so(3)$
in the group $GL(3N)$, with Lie algebra $\gl(3N)$. The outcome in 
the cases $N=1$ and $N=2$ is given in detail.

Section~6 is concerned with an understanding of the position
operator spectrum. To this end, its crucial features are
illustrated in the case $n=2$, both for a 
single particle  two-dimensional ($N=1, D=2$)
model based on the $\gl(1|2)$ representations $V(p)$, and for a two-particle 
one dimensional ($N=2,D=1$) model 
based on the same representations. In both cases a classical limit is recovered by 
taking $p\rightarrow\infty$ and $\hbar\rightarrow0$ in such a way that
$p\hbar\rightarrow C$, for some constant $C$. 

A final example of a WQO model, based on a representation belonging to 
neither the set $W(p)$ nor the set $V(p)$, is given for illustrative
purposes in Section~7. This time the results of Section~2,
that immediately give the allowed energy levels and spatial
coordinates, are augmented by the use of the subalgebra chain
$\gl(3N)\rightarrow \gl(3) \oplus \gl(N)\rightarrow \so(3) \oplus \gl(N)
\rightarrow \so(3)$ to obtain a complete description of
the angular momentum content.

We close with a few brief concluding remarks in Section~8.

Following customary usage in the mathematical physics literature we use in this paper the words
``module'' and ``representation'' more or less interchangeably, and often refer to
``simple module'' as ``irreducible representation.''


\setcounter{equation}{0}
\section{The $\gl(1|n)$ representations} \label{sec:A}%

The Lie superalgebra $\gl(1|n)$ can be defined as the set of all squared 
$(n+1)$-dimensional matrices with rows and columns labelled by indices $i,j=0,1,\ldots,n$. 
As a basis in $\gl(1|n)$ we choose the Weyl matrices $e_{ij}, \; i,j=0,1,\ldots,n$, where 
the odd elements are $\{e_{i0}, e_{0i} | i=1,\ldots,n\}$,
and the remaining elements are even.
The Lie superalgebra bracket is determined by
\begin{equation}
\lb e_{ij}, e_{kl} \rb \equiv e_{ij} e_{kl}-(-1)^{{\deg(e_{ij})
\deg(e_{kl})}}e_{kl}e_{ij} = \delta_{jk} e_{il} - (-1)^{\deg(e_{ij}) \deg(e_{kl})}
\delta_{il} e_{kj}. \label{Weyl}
\end{equation}

Note that $\gl(1|n)_0 = \gl(1) \oplus \gl(n)$, where $\gl(1) =\hbox{span}\{ e_{00}\}$ and
$\gl(n)=\hbox{span}\{ e_{ij}|i,j=1,\ldots,n\}$. For elements $x$ of
$\gl(1|n)$, one defines the supertrace
as $\str(x)=x_{00}-\sum_{j=1}^nx_{jj}$. The Lie superalgebra $\gl(1|n)$ is
not simple, and one can define the simple superalgebra $\ssl(1|n)$
as the subalgebra consisting of elements with supertrace $0$. However,
the representation theory of $\gl(1|n)$ or $\ssl(1|n)$ is essentially
the same (the situation is similar as for the classical Lie
algebras $\gl(n)$ and $\ssl(n)$), and hence we prefer to work with
$\gl(1|n).$ As an ordered basis in the Cartan subalgebra $\h$ of $\gl(1|n)$, we choose
$e_{00}, e_{11}, \ldots, e_{nn}$, and denote by 
\begin{equation}
\epsilon,\delta_1,\ldots,\delta_n \label{ep-de}
\end{equation}
the dual basis in the space $\h^*$ of all linear functionals of $\h$.

The finite-dimensional simple modules of $\gl(1|n)$ are characterized by
their highest weight $\Lambda$. Often, one writes
\begin{equation}
\Lambda = m_{0,n+1}\, \epsilon + \sum_{i=1}^n m_{i,n+1}\, \delta_i .
\end{equation}
Then, the finite-dimensional simple modules $W([m]_{n+1})$ of the Lie 
superalgebra $\gl(1|n)$ are in one-to-one correspondence with the set 
of all complex $n+1$ tuples~\cite{Kac1,Kac2}
\begin{equation}
[m]_{n+1}=[m_{0,n+1}, m_{1,n+1}, \ldots , m_{n,n+1}],  \label{mn+1}
\end{equation}
for which
\begin{equation}
m_{i,n+1}-m_{j,n+1}\in \Z_+, \; \forall i\leq j =1,\ldots, n. \label{cond}
\end{equation}
Within a given $\gl(1|n)$ module $W([m]_{n+1})$ the numbers~(\ref{mn+1}) are fixed.
The possibility of introducing a Gel'fand-Zetlin basis in any finite-dimensional simple 
$\gl(1|n)$ module $W([m]_{n+1})$ stems from the following proposition.
\begin{prop}
Consider the $\gl(1|n)$ module $W([m]_{n+1})$ as a $\gl(n)$ module.
Then $W([m]_{n+1})$ can be represented as a direct sum of simple $\gl(n)$ modules,
\begin{equation}
W([m]_{n+1})=\sum_i \oplus V_i([m]_n), \label{gl(n)}
\end{equation}
where 
\begin{itemize}
\item[I.] All $V_i([m]_n)$ carry inequivalent representations of $\gl(n)$
\begin{equation}
[m]_{n}=[m_{1n}, m_{2n},\ldots , m_{nn}], \; m_{in}-m_{i+1,n}\in\Z_+  .\label{n}
\end{equation}
\item[II.] 
\begin{equation}
 \begin{array}{rl}
1.& m_{in}-m_{i,n+1}=\theta_{i}\in\{0,1\},\quad 1\leq i\leq n,\\
2.& \hbox{if for }\; k\in\{1,\ldots,n\}\; \;
m_{0,n+1}+m_{k,n+1}=k-1, \hbox{ then}\; \theta_k=0.
 \end{array}
\label{cond0}
\end{equation}
\end{itemize}
\end{prop}
The decomposition~(\ref{gl(n)}) and conditions~(\ref{cond0}) follow from the character formula for simple
$\gl(1|n)$ modules~\cite{LJ, JHKR}. If for some $k\in\{1,\ldots,n\}$ the condition $m_{0,n+1}+m_{k,n+1}=k-1$
is satisfied, then the representation is {\em atypical of type~$k$}. Otherwise, it is typical.

A GZ-basis for the $\gl(n)$ module is well known~\cite{GZ}. A set of basis vectors is given 
by a triangular array of
numbers satisfying an integral property and the ``betweenness conditions'':
\begin{equation}
 \left|
\begin{array}{lcllll}
 m_{1n} & \cdots & \cdots & m_{n-1,n} & m_{nn}  \\
 m_{1,n-1} & \cdots & \cdots &  m_{n-1,n-1}  &  \\
\vdots & \qdots & & & \\
m_{11} & & & &
\end{array}
\right).
\label{gln}
\end{equation}
These ``betweenness conditions'' take the form
\begin{equation}
m_{i,j+1}-m_{ij}\in\Z_+\hbox{ and }\; m_{i,j}-m_{i+1,j+1}\in\Z_+,\quad
    1\leq i\leq j\leq n-1 ,
\label{inbetween}
\end{equation}
and they automatically imply the integral property $m_{i,j}-m_{i+1,j}\in\Z_+$ for all meaningful
values of the indices.

Using this GZ-basis for simple $\gl(n)$ modules and Proposition~1 we have 
 
\begin{prop}
The set of vectors 
\begin{equation}
|m) = \left|
\begin{array}{lcllll}
m_{0,n+1} & m_{1,n+1}& \cdots & m_{n-2,n+1} & m_{n-1,n+1} & m_{n,n+1}  \\
& m_{1n} & \cdots & \cdots & m_{n-1,n} & m_{nn}  \\
& m_{1,n-1} & \cdots & \cdots &  m_{n-1,n-1}  &  \\
&\vdots & \qdots & & & \\
&m_{11} & & & &
\end{array}
\right)
\label{m}
\end{equation}
satisfying the conditions
\begin{equation}
 \begin{array}{rl}
1. & m_{i,n+1} \hbox{are fixed and } m_{i,n+1}-m_{j,n+1}\in\Z_+ \quad
    1\leq i\leq j\leq n,\\
2.& m_{in}-m_{i,n+1}=\theta_{i}\in\{0,1\},\quad 1\leq i\leq n,\\
3.& \hbox{if for }\; k\in\{1,\ldots,n\}\; \;
m_{0,n+1}+m_{k,n+1}=k-1, \hbox{then}\; \theta_k=0,  \\
4.& m_{i,j+1}-m_{ij}\in\Z_+\hbox{ and }\; m_{i,j}-m_{i+1,j+1}\in\Z_+,\quad
    1\leq i\leq j\leq n-1.
 \end{array}
\label{cond1}
\end{equation}
constitute a basis in $W([m]_{n+1})$.
\end{prop}
We shall refer to the basis~(\ref{m}) as the GZ-basis for $\gl(1|n)$.
The purpose is now to give the explicit action of a set of $\gl(1|n)$ generators 
on the basis vectors~(\ref{m}). For this purpose,
denote by $|m)_{\pm ij}$ the pattern obtained from $|m)$ by the replacement
$m_{ij} \rightarrow m_{ij}\pm 1$. Then this action is given by:
\begin{eqnarray}
e_{00}|m)&=&\left(m_{0,n+1}-\sum_{j=1}^n \theta_{j}\right)|m); \label{e_00}\\
e_{kk}|m)&=&\left(\sum_{j=1}^k m_{jk}-\sum_{j=1}^{k-1} m_{j,k-1}\right)|m), 
\quad (1\leq k\leq n); \label{e_kk}\\
e_{k-1,k}|m)&=&\sum_{j=1}^{k-1} \left(-
\frac{\prod_{i=1}^{k} (l_{ik}-l_{j,k-1})
\prod_{i=1}^{k-2} (l_{i,k-2}-l_{j,k-1}-1)}{\prod_{i\neq j=1}^{k-1} (l_{i,k-1}-l_{j,k-1})
(l_{i,k-1}-l_{j,k-1}-1) } \right)^{1/2}|m)_{+j,k-1},\quad (2\leq k\leq n); \nn\\[-3mm]
&& \label{ek}\\
e_{k,k-1}|m)&=&\sum_{j=1}^{k-1} \left(-
\frac{\prod_{i=1}^{k} (l_{ik}-l_{j,k-1}+1)
\prod_{i=1}^{k-2} (l_{i,k-2}-l_{j,k-1})}{\prod_{i\neq j=1}^{k-1} (l_{i,k-1}-l_{j,k-1})
(l_{i,k-1}-l_{j,k-1}+1) } \right)^{1/2}|m)_{-j,k-1},\quad (2\leq k\leq n); \nn\\[-3mm]
&& \label{fk}\\
e_{0n}|m)&=&\sum_{i=1}^n \theta_{i}
(-1)^{\theta_{1}+ \ldots +\theta_{i-1} }(l_{i,n+1}+l_{0,n+1}+1)^{1/2}
\left( \frac{\prod_{k=1}^{n-1}  (l_{k,n-1}-l_{i,n+1}-1 )}{\prod_{k\neq i=1}^n (  l_{k,n+1}-l_{i,n+1})}
\right)^{1/2} |m)_{-in}; \nn \\
&& \label{en}\\
e_{n0}|m)&=&\sum_{i=1}^n (1-\theta_{i})
(-1)^{\theta_{1}+ \ldots +\theta_{i-1} }(l_{i,n+1}+l_{0,n+1}+1)^{1/2}\nn\\
 &\times&\left(
\frac{\prod_{k=1}^{n-1} ( l_{k,n-1}-l_{i,n+1} -1)}{\prod_{k\neq i=1}^n (  l_{k,n+1}-l_{i,n+1} )}
\right)^{1/2} |m)_{+in}.\label{fn}
\end{eqnarray}
In all these formulas $l_{ij}=m_{ij}-i$.

In order to deduce~(\ref{e_00})-(\ref{fn}), 
we have used the paper of Palev~\cite{Palev2} and the isomorphism $\varphi$ of $\gl(n|1)$ onto $\gl(1|n)$:
\begin{equation}
\varphi(E_{n,n+1})=e_{0n},\;\; \varphi(E_{n+1,n})=e_{n0},\;\;
\varphi(E_{ij})=e_{ij}, \; i,j=1,\ldots, n,
\end{equation}
where $E_{ij}$ ($i,j=1,\ldots,n+1$) and $e_{ij}$ ($i,j=0,1,\ldots,n$) are the Weyl matrices
of $\gl(n|1)$ and $\gl(1|n)$ respectively.
As compared with the $\gl(n|1)$ case, note that in the right hand 
side of~(\ref{en})-(\ref{fn}) instead of $|m)_{\pm in}$
we have $|m)_{\mp in}$. This is because of the conditions~2 and~3 in Proposition~2. 
Also the atypicality factor $(l_{i,n+1}+l_{0,n+1}+1)$ is different.

For our purposes, it is also necessary to know which of these representations are ``unitary
representations'' with respect to some star condition (or Hermiticity condition) on the 
Lie superalgebra elements. The star condition of relevance is the antilinear anti-involutive
mapping determined by
\begin{equation}
e_{ij}^\dagger = e_{ji}.
\label{star-condition}
\end{equation}
The unitary representations are then those for which there exist a
positive definite inner product $\langle \;| \; \rangle$ (in the representation space $W$)
such that
\begin{equation}
\langle e_{ij}v|w\rangle = \langle v|e_{ji}w\rangle,
\end{equation}
for all $v,w\in W$. These are, in the terminology of~\cite{Gould}, the {\em star type I} representations.
They have been classified in general for $\gl(m|n)$, and for $\gl(1|n)$ the conclusion is
as follows:
\begin{prop}
The representation $W([m]_{n+1})$ is a unitary representation if and only if
\begin{itemize}
\item[(a)] The highest weight is real and 
\begin{equation}
  m_{0,n+1}+m_{n,n+1}-n+1>0.
\end{equation}
In this case, the representation is typical.
\item[(b)] The highest weight is real and there exists a $k\in \{ 1,2,\ldots, n\}$ such that
\begin{equation}
  m_{0,n+1}+m_{k,n+1}=k-1,\quad m_{k,n+1}=m_{k+1,n+1}=\cdots =m_{n,n+1}.
\end{equation}
In this case, the representation is atypical of type~$k$.
\end{itemize}
\end{prop}
For those familiar with the terminology, note that the representations under (b)
essentially correspond to the covariant and contravariant representations of 
$\gl(1|n)$~\cite{JHKT}.
They could also by labelled by $m_{0,n+1}$ and a partition $\lambda$.

The representations $W([m]_{n+1})$ are unitary if the conditions of 
Proposition~3 are satisfied, for the inner product corresponding to
orthonormal GZ basis vectors, i.e.
\begin{equation}
\langle |m')\,|\; |m) \rangle \equiv (m'|m) = \delta_{m,m'},
\end{equation}
under the action~(\ref{e_00})-(\ref{fn}).

We end this section with a technical result. 
In principle, it is sufficient to have the action of the $\gl(1|n)$ generators
$e_{kk}$ ($k=0,\ldots,n$), $e_{k-1,k}$, $e_{k,k-1}$ ($k=1,\ldots,n$), and
$e_{0n}$, $e_{n0}$ as given in~(\ref{e_00})-(\ref{fn}), in order to compute the action of
any element $e_{ij}$ on the GZ basis vectors~(\ref{m}). 
For our later application, it is useful to know the explicit action of the odd elements
$e_{0j}$ and $e_{j0}$ of $\gl(1|n)$. It is found that this is given by the
following complicated formulae:
\begin{eqnarray}
e_{0j}|m)&=&\sum_{i_n=1}^n\sum_{i_{n-1}=1}^{n-1}\ldots \sum_{i_j=1}^j
\theta_{i_n}(-1)^{\theta_1+\ldots +\theta_{i_n-1}}(l_{i_n,n+1}+l_{0,n+1}+1)^{1/2}\nn\\
&\times&
\prod_{r=j+1}^nS(i_r,i_{r-1})
\left( \frac{\prod_{k\neq i_{r-1}=1}^{r-1}  
(l_{k,r-1}-l_{i_r,r} )\prod_{k\neq i_r=1}^r(l_{kr}-l_{i_{r-1},r-1}+1)
 }{ \prod_{k\neq i_r=1}^r (l_{kr}-l_{i_r,r})\prod_{k\neq i_{r-1}=1}^{r-1}(l_{k,r-1}
 -l_{i_{r-1},r-1}+1)}
\right)^{1/2} \label{e0j}\\
&\times &
\left({\prod_{k\neq i_n=1}^n } \frac{(l_{kn}-l_{i_n,n} )}{(  l_{k,n+1}-l_{i_n,n+1})}
\right)^{1/2}
\left( \frac{\prod_{k=1}^{j-1} (l_{k,j-1}-l_{i_j,j} )}{\prod_{k\neq i_j=1}^j(  l_{kj}-l_{i_j,j})}
\right)^{1/2}
|m)_{-i_n,n;-i_{n-1},n-1;\ldots ;-i_j,j}\nn
\end{eqnarray}

\begin{eqnarray}
e_{j0}|m)&=&\sum_{i_n=1}^n\sum_{i_{n-1}=1}^{n-1}\ldots \sum_{i_j=1}^{j}
(1-\theta_{i_n})(-1)^{\theta_1+\ldots +\theta_{i_n-1}}(l_{i_n,n+1}+l_{0,n+1}+1)^{1/2}\nn\\
&\times&
\prod_{r=j+1}^nS(i_r,i_{r-1})
\left( \frac{\prod_{k\neq i_{r-1}=1}^{r-1}  
(l_{k,r-1}-l_{i_r,r}-1 )\prod_{k\neq i_r=1}^r(l_{kr}-l_{i_{r-1},r-1})
 }{  \prod_{k\neq i_r=1}^r (l_{kr}-l_{i_r,r})\prod_{k\neq i_{r-1}=1}^{r-1}(l_{k,r-1}
 -l_{i_{r-1},r-1}-1)}
\right)^{1/2} \label{ej0}\\
&\times &
\left({\prod_{k\neq i_n=1}^n } \frac{(l_{kn}-l_{i_n,n} )}{(  l_{k,n+1}-l_{i_n,n+1})}
\right)^{1/2}
\left( \frac{\prod_{k=1}^{j-1} (l_{k,j-1}-l_{i_j,j}-1 )}{\prod_{k\neq i_j=1}^j(  l_{kj}-l_{i_j,j})}
\right)^{1/2}
|m)_{+i_n,n;+i_{n-1},n-1;\ldots ;+i_j,j},\nn
\end{eqnarray}
where $j=1,\ldots ,n$, each symbol $\pm i_k,k$ attached as a subscript to $|m)$ indicates a
replacement $m_{i_k,k}\rightarrow m_{i_k,k}\pm 1$, and 
\begin{eqnarray}
&&S(k,l) = \left\{ \begin{array}{lll}
 {\;\;1} & \hbox{for} & k\leq l  \\ 
 {-1} & \hbox{for} & k>l .
 \end{array}\right.
\end{eqnarray}


\setcounter{equation}{0}
\section{A special class of representations} \label{sec:covariant}%

One interesting class of representations of $\gl(1|n)$ is that with 
$[m]_{n+1}=[p,0,\ldots,0]$, i.e.\ with highest weight $\Lambda = p \epsilon$,
where $p\in\Z_+$.
These are covariant representations labelled by the partition $(p)$. 
The representation space $W([p,0,\ldots,0])$ is simply denoted by $W(p)$.
The GZ basis vectors of $W(p)$ can be denoted by $|p;\varphi_1,\ldots,\varphi_n\rangle$, where
the relation to the GZ labels is determined by
\begin{equation}
\varphi_i= \sum_{j=1}^i m_{ji} - \sum_{j=1}^{i-1} m_{j,i-1}.
\end{equation}
The constraints for the GZ labels lead to: $\varphi_i\in\{0,1\}$ and 
$\sum_{i=1}^n \varphi_i \leq \min(p,n)$.
The representations $W(p)$ and the basis vectors $|p;\varphi_1,\ldots,\varphi_n\rangle$
have been constructed by means of Fock space techniques,
and the action of the $\gl(1|n)$ generators is very simple, see~\cite{Palev3}.
This class is referred to as the class of Fock representations.

In this paper, another special class of representations will be of importance,
namely those with $[m]_{n+1}=[1,p-1,0,\ldots,0]$, i.e.\ with highest weight 
$\Lambda = \epsilon+(p-1)\delta_1$, where $p\in\Z_+$. These are again 
covariant irreducible representations, labelled this time by the
partition $(1,1,\ldots,1)=(1^p)$.
The representation space $W([1,p-1,0,\ldots,0])$ will be denoted by $V(p)$.
We shall assume that $p>1$, since for $p=1$ it is actually a Fock representation.
Note that $V(p)$ is atypical of type~2, and by Proposition~3 it is also unitary.

One can now apply the general GZ-basis construction of the previous section
to this special class.
The GZ basis vectors of $V(p)$ are of the form 
\begin{equation}
\left|
\begin{array}{lcllll}
1 & p-1 & 0 & \cdots & 0 &0   \\
& q_n & 0 & \cdots & 0 & 0  \\
& q_{n-1} & 0 & \cdots &  0  &  \\
&\vdots & \qdots & & & \\
&q_2 & 0& & &\\
&q_{1} & & & &
\end{array}
\right)
\label{q}
\end{equation}
with $q_n=p-\theta \geq q_{n-1} \geq \cdots \geq q_1 \geq 0$, where $\theta =0,1$
(note that $\theta=1-\theta_1$ in the description of~(\ref{cond1})).
Since most of the entries in the GZ-array are zero, it will be more convenient to denote 
this vector by
\begin{equation}
| \theta; q_n,q_{n-1},\ldots,q_1 \rangle \qquad 
(q_n=p-\theta \geq q_{n-1} \geq \cdots \geq q_1 \geq q_0 \equiv 0).
\end{equation}
The actions~(\ref{e_00})-(\ref{fn}) and~(\ref{e0j})-(\ref{ej0}) imply that
\begin{eqnarray}
&& e_{00} |\theta; q_n,q_{n-1},\ldots,q_1 \rangle = \theta |\theta; q_n,q_{n-1},\ldots,q_1 \rangle; \\
&& e_{kk} |\theta; q_n,q_{n-1},\ldots,q_1 \rangle = (q_k-q_{k-1}) |\theta; q_n,q_{n-1},\ldots,q_1 \rangle,
\qquad (1\leq k\leq n);\\
&& e_{k,k-1} |\theta; q_n,q_{n-1},\ldots,q_1 \rangle =\sqrt{(q_k-q_{k-1}+1)(q_{k-1}-q_{k-2})} \nn\\
&& \qquad\times\  |\theta; q_n,q_{n-1},\ldots,q_k,q_{k-1}-1,\ldots,q_1 \rangle, \qquad (2\leq k \leq n);\\
&& e_{k-1,k} |\theta; q_n,q_{n-1},\ldots,q_1 \rangle =\sqrt{(q_k-q_{k-1})(q_{k-1}-q_{k-2}+1)} \nn\\
&& \qquad \times\  |\theta; q_n,q_{n-1},\ldots,q_k,q_{k-1}+1,\ldots,q_1 \rangle, \qquad (2\leq k \leq n);\\
&& e_{k0} |\theta; q_n,q_{n-1},\ldots,q_1 \rangle =\theta \sqrt{q_k-q_{k-1}+1} \nn\\
&& \qquad\times\  |1-\theta; q_n+1,q_{n-1}+1,\ldots,q_k+1,q_{k-1},\ldots,q_1 \rangle,\qquad(1\leq k\leq n);\\
&& e_{0k} |\theta; q_n,q_{n-1},\ldots,q_1 \rangle =(1-\theta)\sqrt{q_k-q_{k-1}} \nn\\
&& \qquad \times\  |1-\theta; q_n-1,q_{n-1}-1,\ldots,q_k-1,q_{k-1},\ldots,q_1 \rangle, \qquad(1\leq k\leq n).
\end{eqnarray}
These formulas become even simpler in yet another notation for the basis vectors.
Let us put
\begin{equation}
r_n=q_n-q_{n-1}, r_{n-1}=q_{n-1}-q_{n-2}, \ldots, r_{2}=q_2-q_1, r_1=q_1,
\end{equation}
then all $r_i$ are nonnegative integers, with $|r|=r_1+\cdots+r_{n}=q_n$.
For the vectors with $\theta=1$, $q_n=p-1$, so $|r|=p-1$.
For the vectors with $\theta=0$, $q_n=p$, so $|r|=p$.
Thus all vectors of $V(p)$ are described by:
\begin{equation}
v(\theta;r)\equiv v(\theta;r_1,r_2, \ldots,r_n),\qquad
\theta\in\{0,1\},\ r_i\in\{0,1,2,\ldots\},\ \hbox{ and } \theta+r_1+\cdots+r_n= p.
\label{newbasis}
\end{equation}
In this notation the highest weight vector is $v(1;p-1,0,\ldots,0)$.
The action of the $\gl(1|n)$ generators on the new basis~(\ref{newbasis}) is now given by:
\begin{eqnarray}
&& e_{00} v(\theta; r) = \theta v(\theta; r);\\
&& e_{kk} v(\theta; r) = r_k v(\theta; r),\qquad (1\leq k\leq n);\\
&& e_{k,k-1} v(\theta; r) =\sqrt{r_{k-1}(r_k+1)}\
  v(\theta; r_1,\ldots,r_{k-1}-1, r_k+1,\ldots,r_{n}),\quad(2\leq k\leq n);\\
&& e_{k-1,k} v(\theta; r) =\sqrt{(r_{k-1}+1)r_k}\
  v(\theta; r_1,\ldots,r_{k-1}+1, r_k-1,\ldots,r_{n}), \quad(2\leq k\leq n);\\
&& e_{k0} v(\theta; r) =\theta \sqrt{r_k+1}\ v(1-\theta; r_1,\ldots,r_k+1,\ldots,r_n), \qquad (1\leq k\leq n);\\
&& e_{0k} v(\theta; r) =(1-\theta)\sqrt{r_k}\ v(1-\theta; r_1,\ldots,r_k-1,\ldots,r_n),
\qquad (1\leq k\leq n).
\end{eqnarray}
With respect to the inner product 
\begin{equation}
\langle v(\theta;r) | v(\theta';r') \rangle = \delta_{\theta,\theta'}\delta_{r,r'},
\end{equation}
the representation $V(p)$ is unitary for the star condition~(\ref{star-condition}).


\setcounter{equation}{0}
\section{The $N$-particle $D$-dimensional WQO} \label{sec:WQO}%

\subsection{The WQO operators}

The rest of this paper is devoted to new solutions of the WQO and their physical interpretation.
Let us briefly recall the context of WQOs~\cite{Palev1}-\cite{SJ}. 
Let $\hat{H}$ be the Hamiltonian of an $N$-particle $D$-dimensional
harmonic oscillator, that is
\begin{equation}
\hat{H}=\sum_{\alpha=1}^{N} \Big( \frac{\hbP_\alpha^2}{2m}
+ \frac{m\omega^2}{2} {\hbR}_\a^2 \Big), \label{H}
\end{equation}
with $\hbP_\alpha$ and $\hbR_\alpha$ $D$-dimensional vector operators
corresponding to the momentum and position of the particle~$\alpha$ ($\alpha=1,2,\ldots,N$), $m$
the mass and $\omega$ the frequency of each oscillator.
We proceed to view this oscillator as a Wigner quantum system: this means
that the canonical commutation relations are not required, but are replaced 
by compatibility conditions between Hamilton's equations and the
Heisenberg equations. In other words, Hamilton's equations
\begin{equation}
    {\dot{\hbP}}_\a=-m\omega^2\hbR_\a, \ \ {\dot{\hbR}}_\a = \frac{1}{m}\hbP_\a
    \ ~ {\rm for} ~\ \a=1,2,\ldots,N,
     \label{Ham}
\end{equation}
and the Heisenberg equations
\begin{equation}
     {\dot{\hbP}}_\a = \frac{i}{\hbar}[\hat{H},\hbP_\a], \ \
     {\dot{\hbR}}_\a = \frac{i}{\hbar}[\hat{H},\hbR_\a]
     \ ~ {\rm for} ~ \ \a=1,2\ldots,N,
     \label{Heis}
\end{equation}
should be identical as operator equations. These compatibility conditions are
such that
\begin{equation}
   [\hat{H},\hbP_\a]=i\hbar m \omega^2\hbR_\a ,\ \
   [\hat{H},\hbR_\a]=-\frac{i\hbar}{m}\hbP_\a
    \ ~ {\rm for} ~ \ \a=1,2,\ldots,N.
     \label{comp}
\end{equation}
To make the connection with $\ssl(1|DN)$ we write the operators
$\hbP_\a$ and $\hbR_\a$
for $\a=1,2,\ldots,N$ in terms of new operators:
\begin{equation}
A_{D(\alpha -1) +k}^\pm= \sqrt{\frac{(DN-1)m \omega}{4\hbar}}
 \hR_{\a k} \pm i
  \sqrt { \frac{(DN-1)}{4m \omega \hbar}} \hP_{\a k}, \quad k=1,\ldots,D.
 \label{A}
\end{equation}
The Hamiltonian $\hat{H}$ of~(\ref{H}) and the
compatibility conditions~(\ref{comp}) take the form:
\begin{equation}
     \hat{H}= \frac{\omega \hbar}{DN-1} \sum_{j=1}^{DN} \{A_j^+, A_j^-\},
     \label{HAA}
\end{equation}
\begin{equation}
\sum_{j=1}^{DN}  [ \{A_{j}^+,A_{j}^- \},A_{i}^\pm]
=\mp (DN-1)A_{ i}^\pm , \quad i,j=1,2,\ldots , DN.
\label{comp1}
\end{equation}
As a solution to~(\ref{comp1}) one can choose operators $A_{ i}^\pm$ that satisfy the
following triple relations:
\begin{eqnarray}
&& [\{A_{i}^+,A_{j}^-\},A_{k}^+]=
\delta_{jk}A_{i}^+
-\delta_{ij}A_{k}^+,  \nn\\
&& [\{A_{ i}^+,A_{ j}^-\},A_{ k}^-]=
-\delta_{ik}A_{j}^-
+\delta_{ij}A_{k}^-, \label{sl}\\
&& \{A_{i}^+,A_{ j}^+\}=
\{A_{ i}^-,A_{ j}^-\}=0. \nn
\end{eqnarray}

\begin{prop}
The operators $A_{j}^\pm$, for $j=1,2,\ldots,DN$,
are odd elements generating the Lie superalgebra
$\ssl(1|DN)$.
\end{prop}
This means that one can identify the operators $A^+_j$ and $A^-_j$ with the $\ssl(1|DN)$ 
generators $e_{j0}$ and $e_{0j}$ respectively:
\begin{equation}
A_j^+=e_{j0},\qquad A_j^-=e_{0j}.
\end{equation}
With this identification it should be noted that
\begin{equation}
\hat{H}=\frac{\hbar\omega}{(DN-1)} \sum_{k=1}^{DN} (e_{00}+e_{kk}) =
\frac{\hbar\omega}{(DN-1)} \left( DN\,e_{00}+ \sum_{k=1}^{DN} e_{kk} \right).
\label{Hee}
\end{equation}

To end this subsection, let us also recall that in the case $D=3$ it is possible to
introduce the angular momentum operator of each particle $\alpha$.
These single particle angular momentum operators
$\hJ_{\a j}$ are defined by~\cite{Palev5}
\begin{equation}
      \hJ_{\a j} = -\frac{3N-1}{2\hbar} \sum_{k,l=1}^3
    \ \epsilon_{jkl} \{\hR_{\a k},\hP_{\a l}\}
      \ \ \a=1,2,\ldots,N,\ \ j=1,2,3,
      \label{MRP}
\end{equation}
and take the following form:
\begin{equation}
      \hJ_{\a j} = -i \sum_{k,l=1}^3 \epsilon_{jkl}
   \{ A_{3(\a-1)+k}^+,A_{3(\a-1)+ l}^-\}.
      \label{MA}
\end{equation}
In terms of these operators the three components of the total angular
momentum operator $\hat{\J}$ are given by
\begin{equation}
\hJ_j=\sum_{\a=1}^N \hJ_{\a j},\quad j=1,2,3. \label{M}
\end{equation}
It is straightforward to verify that with respect to this choice
of angular momentum operator $\hat{\J}$ the operators $\hbR_\a$, $\hbP_\a$,
$\hat{\J}_\a$ and $\hat{\J}$ all transform as $3$-vectors.

\subsection{The WQO representations}

The Hilbert space (state space) of the WQO is a representation space $W$ of the
Lie superalgebra $\gl(1|DN)$ in such a way that the ``observables'' (the position and
momentum operators $\hat R_{\alpha,k}$ and $\hat P_{\alpha,k}$) are Hermitian operators.
This means that $(A_j^\pm )^\dagger = A_j^\mp$, or $e_{j0}^\dagger=e_{0j}$. 
In other words, the state spaces are unitary representations of $\gl(1|DN)$, in the
sense of Proposition~3.
Since all ``physical operators'' in this WQO model are expressed in terms of
the $A_j^\pm$, and their actions are known by~(\ref{e0j})-(\ref{ej0}),
all the relevant physical properties can in principle be deduced.

In previous papers, an investigation was made of the physical properties of the
$\gl(1|DN)$ solutions in the Fock representation spaces $W(p)$, see~\cite{Palev1, K1}
 for $D=3$, $N=1$ 
(the single particle 3-dimensional WQO) and~\cite{Palev5, K2}
for $D=3$ and $N$ arbitrary (the last case corresponding to a superposition
of $N$ single particle 3-dimensional WQOs).
The most striking properties are as follows:
the energy of each particle has at most four different eigenvalues (equidistant energy levels);
the geometry is non-commutative, in the sense that coordinate operators do not commute;
the position and momentum operators have discrete spectra.
In this paper, we shall consider more general solutions corresponding to
arbitrary unitary representations of $\gl(1|DN)$, and to the special class
of representations $V(p)$.

Some of these properties will be dealt with
in general (arbitrary $N$ and $D$, arbitrary unitary representations); 
for some others it will be useful to take particular values of $N$ and/or $D$,
or to restrict oneself to specific classes of unitary representations.
In the general case, all unitary representations of $\gl(1|DN)$ are of relevance.
In order to simplify notation, we shall use the abbreviation
\begin{equation}
n=DN
\label{n=DN}
\end{equation}
throughout this section.

\subsubsection{Energy spectrum}

The Hamiltonian $\hat{H}$ is diagonal in the GZ-basis,
i.e.\ the basis vectors $|m)$ are stationary states
of the system. This follows from the form of $\hat{H}$ given in~(\ref{Hee})
and the action of $e_{00}$ and $e_{kk}$ given in~(\ref{e_00}) and~(\ref{e_kk}).
In each space $W([m]_{n+1})$ there is a finite number of equally
spaced energy levels, with spacing $\hbar\omega$: 
\begin{equation}
     \hat{H} |m) = E_q |m) 
\end{equation}
with
\begin{equation}
  E_q={\hbar\omega}\left( \frac{n m_{0,n+1}+m_{1,n+1}+\ldots+m_{n,n+1}}{n-1}-q\right),
\end{equation}
where  $q=\sum_{j=1}^{n}\theta_j$.
For typical representations, $q$ takes the values $0,1,\ldots,n$.
For a unitary representation $W([m]_{n+1})$ atypical of type~$k$, $q$ takes the values $0,1,\ldots, k-1$.
Recall that in that case, the representation labels satisfy
\begin{equation}
     m_{0,n+1}+m_{k,n+1}=k-1, \qquad m_{k,n+1}=m_{k+1,n+1}=\cdots=m_{n,n+1}.
\label{Eq}
\end{equation}
So for a typical representation, there are $n+1$ equidistant energy levels; 
for an atypical representation of type~$k$
($k=1,2,\ldots,n$) there are $k$ equidistant energy levels. 
The degeneracy of these levels is high, and can be determined explicitly from the GZ-basis labels,
or equivalently from the dimensions of the $\gl(n)$ representations in the decomposition
of $W([m]_{n+1})$.

For example, for the special class of representations $V(p)$, atypical of type $2$, 
there are only two distinct energy levels:
\begin{eqnarray*}
&&E_0={\hbar\omega}(\frac{p}{n-1})\hbox{ with degeneracy } 
\left(\begin{array}{c}{p+n-1}\\{n-1}\end{array}\right),\\
&&E_1={\hbar\omega}(\frac{p}{n-1}+1)\hbox{ with degeneracy } 
\left(\begin{array}{c}{p+n-2}\\{n-1}\end{array}\right).
\end{eqnarray*}

\subsubsection{Position and momentum operators}

The position operators $\hat R_{\a k}$ ($\a =1,\ldots, N$, $k=1,\ldots,D$)
of the oscillating particles do not commute with each other
\begin{equation}
[\hat R_{\a i},\hat R_{\b j}]\ne 0 \quad \hbox{ for } \quad \a i\ne \b j.
\label{3.3}
\end{equation}
Similarly
\begin{equation}
[\hat P_{\a i},\hat P_{\b j}]\ne 0 \quad \hbox{ for } \quad \a i\ne \b j.
\label{3.4}
\end{equation}
These relations imply that the WQO belongs to the class of models of non-commutative
quantum oscillators~\cite{NP}-\cite{MM}, or to theories with non-commutative 
geometry~\cite{HS, C}. 
In general, the action of $[\hat R_{\a i},\hat R_{\b j}]$ is difficult to describe and interpret,
even when acting on GZ basis vectors of particular unitary representations.
For example, for the representations $V(p)$, one finds:
\begin{eqnarray}
&& [\hat R_{\a i},\hat R_{\b j}] v(\theta;r) = \frac{\hbar}{(n-1)m\omega} \Bigl( 
(-1)^\theta \sqrt{r_l(r_k+1)} v(\theta;r_1,\ldots,r_k+1,\ldots,r_l-1,\ldots,r_{n}) \nn\\
&& \qquad - (-1)^\theta \sqrt{r_k(r_l+1)} v(\theta;r_1,\ldots,r_k-1,\ldots,r_l+1,\ldots,r_{n})\Bigr),
\end{eqnarray}
where $k=D(\a-1)+i$ and $l=D(\b-1)+j$ (and it is assumed that $k<l$).

However, the squares of the components of position and momentum operators commute:
\begin{equation}
[\hat R_{\a i}^2,\hat R_{\b j}^2]=[\hat P_{\a i}^2,\hat P_{\b j}^2]= 0 \quad \hbox{ for } \quad \a i\ne \b j .
\label{3.5}
\end{equation}
Furthermore, the GZ basis states $|m)$ are eigenstates of these operators,
\begin{eqnarray}
\hat R_{\a i}^2 |m)&=& \frac{\hbar}{(n-1)m\omega} (m_{0,n+1}+\ldots+m_{n,n+1}
-m_{1,n}-\cdots-m_{n,n}\nn\\
&& +m_{1,k}+\cdots +m_{k,k}-m_{1,k-1}-\cdots -m_{k-1,k-1})|m),
\label{actionR2}
\end{eqnarray}
where $k=D(\a-1)+i$.
Thus the spectrum of the position operator component $\hat R_{\a i}$ is given by 
the set of values
\begin{equation}
\pm \sqrt{ \frac{\hbar}{(n-1)m\omega} (\sum_{j=0}^{n}
 m_{j,n+1}-\sum_{j=1}^{n}m_{j,n}+\sum_{j=1}^{k}m_{j,k}
 -\sum_{j=1}^{k-1}m_{j,k-1}}),
  \label{rev}
\end{equation}
where $k=D(\a-1)+i$ and the internal labels $m_{j,l}$ take all possible values 
allowed by~(\ref{cond1}).
For the special representations $V(p)$, (\ref{actionR2}) becomes
\begin{equation}
\hat R_{\a i}^2 v(\theta;r) = \frac{\hbar}{(n-1)m\omega}(r_k+\theta) v(\theta;r),\qquad k=D(\a-1)+i.
\label{R2onVp}
\end{equation}
This gives rise to a simple spectrum of the operators $\hat R_{\a i}$ in these representations.

\setcounter{equation}{0}
\section{Angular momentum content for $V(p)$}

As mentioned before, the case $D=3$ allows us to introduce angular momentum operators, 
see~(\ref{MRP})-(\ref{M}).
It is easy to verify that in general the stationary states $|m)$ are not 
eigenstates of $\hat{J}_{\a 3}$, $\hat{\J}_\a$, $\hat{J}_3$
and $\hat{\J}^2$. Thus in order to determine the
possible values of the total angular momentum $j$ for the $N$-particle 3-dimensional WQO,
it is best to use group theoretical methods rather than the explicit actions 
of these operators on basis states.
The components $\hat{J}_k$ in~(\ref{M}) are the generators of an $\so(3)$ subalgebra of
$\gl(1|n)=\gl(1|3N)$, that may be identified by means of the following chain of subalgebras:
\begin{equation}
    \gl(1|3N)\rightarrow \gl(1)\oplus \gl(3N)
\rightarrow \gl(1) \oplus \gl(3) \oplus \gl(N)
\rightarrow \gl(1) \oplus \so(3) \oplus \gl(N)
\rightarrow \gl(1) \oplus \so(3).
     \label{gl13n}
\end{equation} 
The first step in the branching rule, from $\gl(1|3N)$ to $\gl(1)\oplus \gl(3N)$, is
just determined by the GZ-pattern~(\ref{m}), where the $\gl(1)$ value is given by~(\ref{e_00}).
The branching rule for $\gl(3N)\rightarrow \gl(3)\oplus \gl(N)$
required in the second step and that for $\gl(3)\rightarrow \so(3)$ required in the third
step are both rather well known and have been implemented for example in
SCHUR~\cite{Schur}. Since they involve coefficients for which there is
no known general formula, we content ourselves with giving the results explicitly just
for the special representations $V(p)$.

Before doing this we should remark that the embedding of $\so(3)$ in $\gl(3N)$ is such that
the defining $3N$-dimensional representation $V^{\{1\}}_{\gl(3N)}$ of $\gl(3N)$ decomposes 
into a direct sum of $N$ copies of the defining $3$-dimensional representation $V^{1}_{\so(3)}$
of $\so(3)$, so that all states are of angular momentum $j=1$. The $p$th-fold tensor powers
of $V^{\{1\}}_{\gl(3N)}$ therefore contain only states whose angular momentum $j$ is bounded
by $p$. This includes all covariant irreducible representations $V^{\{\pi\}}_{\gl(3N)}$
where $\pi$ is any partition of weight $p$, including the special case $V^{\{p\}}_{\gl(3N)}$
of particular interest here.
 
With respect to the first step of (\ref{gl13n}), 
the representation $V(p)$ of $\gl(1|3N)$ decomposes into just two symmetric 
irreducible representations of $\gl(3N)$:
\begin{equation}
V(p) \rightarrow V^{0}_{\gl(1)} \otimes V^{\{p\}}_{\gl(3N)} + V^{1}_{\gl(1)} \otimes V^{\{p-1\}}_{\gl(3N)},
\label{step1}
\end{equation}
labelled by the partitions $\{p\}$ and $\{p-1\}$ respectively.
For symmetric representations labelled by a positive integer $p$, the branching from
$\gl(3N)$ to $\gl(3)\oplus \gl(N)$ is determined by the branching rule~\cite{K3}: %
\begin{equation}
V^{\{p\}}_{\gl(3N)} \rightarrow \sum_{\lambda, |\lambda|=p} V^{\{\lambda\}}_{\gl(3)} \otimes V^{\{\lambda\}}_{\gl(N)},
\label{step2}
\end{equation}
where the sum is over all partitions $\{\lambda\}$ with $\lambda_1+\lambda_2+\cdots =p$. 
Furthermore, the labelling of representations of $\gl(3)$ and $\gl(N)$ requires the extra
condition that the length of $\{\lambda\}$ should be at most $\min(3,N)$.

In the case $N=1$, this step is absent, and for the branching $\gl(3)\rightarrow \so(3)$ we have immediately
\begin{equation}
V^{\{p\}}_{\gl(3)} \rightarrow V^{p}_{\so(3)} + V^{p-2}_{\so(3)}+\cdots V^{1\ \hbox{\tiny or }0}_{\so(3)}.
\end{equation}
Thus, using~(\ref{step1}), the total angular momentum values are given by $j=0,1,\ldots,p$, with no
multiplicities, so that
\begin{equation}
V(p) \rightarrow \sum_{j=0}^p V^{j}_{\so(3)}.
\label{VpN1}
\end{equation}

In the case $N\geq2$, it might appear that the second step, (\ref{step2}),
cannot be avoided. However, if our intention is simply to identify all
possible angular momentum states $j$ and their multiplicities
that arise for $V^{\{p\}}_{\gl(3N)}$, then this may be accomplished
by exploiting some classical results from invariant theory. 
For this purpose, we shall briefly use the Lie group embedding $GL(3N)\supset SO(3)$
rather than the Lie algebra embedding $\gl(3N)\supset \so(3)$.
For the current class of representations the branching rules associated
with these two embeddings are identical.

For any compact continuous subgroup $G$ of $GL(n)$,
with group elements $g$ and Haar measure $d\mu(g)$, Molien's function~\cite{Molien}
takes the form~\cite{Sturmfels}
\begin{equation}
 M(P) = \int \frac{d\mu(g)}{\det(I-P\,g)},
\label{Mol}
\end{equation}
where $I$ is the unit $n\times n$ matrix. The significance of this
function in invariant theory is that when expanded in the form:
\begin{equation}
  M(P) = \sum_{p=0}^\infty \, n_p\, P^p\,,  
\label{MP}
\end{equation}
the expansion coefficient $n_p$ is the number of linearly independent 
invariants of $G$ that are homogeneous polynomials in the matrix elements
of $g$ of degree $p$. Equivalently, $n_p$ is the multiplicity of 
the trivial 1-dimensional identity representation of $G$ 
in the restriction of $V^{\{p\}}_{GL(n)}$ to $G$. More generally~\cite{RRS},
if the irreducible representation $V^j_{G}$, specified by some
label $j$, has character $\chi^j(g)$, with complex
conjugate $\chi^{j\ast}(g)$, then 
\begin{equation}
M^j(P)=\int \frac{\chi^{j\ast}(g)\, d\mu(g)}{\det(I-P\,g)}
\label{MolChi}
\end{equation}
has an expansion of the form
\begin{equation}
  M^j(P) = \sum_{p=0}^\infty\, n_{p,j}\, P^p\,.
\label{MPj}
\end{equation}
where each coefficient $n_{p,j}$ is the multiplicity of 
the irreducible representation $V^j_{G}$ of $G$ 
in the restriction of $V^{\{p\}}_{GL(n)}$ to $G$. 

Specialising this to the case $n=3N$ and $G=SO(3)$ we immediately have a
means of calculating the angular momentum content of the
representation $V^{\{p\}}_{GL(3N)}$
as has been done in the case $N=2$ by Raychev {\it et al}~\cite{RRS}.
To this end it should be noted that
each element $g$ of $SO(3)\subset GL(3N)$ arises as the $N$th-fold
tensor power of a $3\times 3$ matrix which can itself be diagonalised,
with diagonal elements $(z,1,z^{-1})$ where $z=\exp(i\phi)$ for some real
$\phi$. With this parametrisation, taking into account the diagonalisation 
process, the relevant Haar measure is given by~\cite{RRS}
\begin{equation}
      d\mu(g)= \frac{1}{2}\ (1-\cos\phi) \, d\phi\,,
\label{mu-g}
\end{equation} 
with $0\leq\phi<2\pi$.
Moreover, the diagonal form of $g$ is such that
\begin{equation}
      \det(I-P\,g)=\left( (1-P\,z)(1-P)(1-P\,z^{-1})\right)^N. 
\label{det-Pz}
\end{equation} 
In addition it is well known that the character of the irreducible
representation $V^j_{SO(3)}$ of $SO(3)$ is given by
\begin{equation}
     \chi^j(g) = \frac { z^{j+\frac12}-z^{-j-\frac12} }
         { z^{\frac12}-z^{-\frac12} }\,,
\label{chi-j}
\end{equation}
where it is to be noticed that $\chi^{j\ast}(g) =\chi^j(g)$.
In our case, only integer (rather than half-integer) values of $j$ arise.
This allows us to introduce a generating function for $SO(3)$ characters
as follows:
\begin{eqnarray}
\hbox{char}(J,g)&=& \sum_{j=0}^\infty \, J^j\, \chi^j(g) 
= \sum_{j=0}^\infty\,  \frac{ (J\,z)^j\, z^{\frac12}-(J\,z^{-1})^j\, z^{-\frac12} }
        { z^{\frac12}-z^{-\frac12} }\cr
&=& \sum_{j=0}^\infty \left( \frac {(J\,z)^j}{1-z^{-1}} - \frac{(J\,z^{-1})^j}{z-1}\right)\cr
&=& \frac {1}{(1-J\,z)(1-z^{-1})}-\frac{1}{(1-J\,z^{-1})(z-1)}.
\label{charJ}
\end{eqnarray}
Hence, we have
\begin{eqnarray}
&&M(P,J)_N=\sum_{j=0}^\infty M^j(P)\,J^j
= \int \frac{ \hbox{char}(J,g^{-1})\, d\mu(g)}{\det(I-P\,g)} \cr
&&\ = \oint_{|z|=1} \left(\frac {(1-z^{-1})^{-1} }{(1-J\,z)}
               -\frac{(z-1)^{-1}}{(1-J\,z^{-1}) } \right)
\frac { \left( 1-\frac12(z+z^{-1})\right) } 
      { \left( (1-P\,z)(1-P)(1-P\,z^{-1})\right)^N     }
\frac{dz}{4\pi iz}\,,    
\label{MPJ}
\end{eqnarray}
where the fact that $z=\exp(i\phi)$ has allowed the integral over 
$\phi$ to be re-written as an integral around the unit circle in the 
complex $z$-plane.

In fact, in view of the rather simple dependence on $N$, we can go even further 
and introduce the master generating function:
\begin{eqnarray}
\hskip-1cm
&&M(P,J,{\cal N})=\sum_{N=0}^\infty M(P,J)_N\,{\cal N}^N
=\sum_{p,j,N=0}^\infty n_{p,j,N}\, P^p\,J^j\,{\cal N}^N \cr
\hskip-1cm
&&\ = \oint_{|z|=1} \left(\frac {(1-z^{-1})^{-1} }{(1-J\,z)}
               -\frac{(z-1)^{-1}}{(1-J\,z^{-1}) } \right)
\left( \frac{\left( 1-\frac12(z+z^{-1})\right)} 
        {1-{\cal N}/((1-P\,z)(1-P)(1-P\,z^{-1}))} \right) \frac{dz}{4\pi  iz}\,,    
\label{MPJN}
\end{eqnarray}
where $n_{p,j,N}$ is the multiplicity of the angular momentum state $j$ in the
irreducible representation $V^{\{p\}}_{GL(3N)}$ of $GL(3N)$, or equivalently in
the representation $V^{\{p\}}_{\gl(3N)}$ of $\gl(3N)$. This integral
may be evaluated by determining the residues of the integrand at all poles within 
the unit circle. The result is
\begin{eqnarray}
&&M(P,J,{\cal N}) = 
1 + \frac12\, \frac{{\cal N}(1-J)} {J{\cal N}-(1-P)(1-JP)(J-P)}\cr
&&{\phantom{M(P,J,{\cal N}) =}}\  + \frac12\, \frac{{\cal N}(1+J)} {(J{\cal N}-(1-P)(1-JP)(J-P)}\,
    \sqrt{ \frac {{\cal N}-(1-P)^3 } {{\cal N}-(1-P)(1+P)^2   }  }
    \label{masterGF}
\end{eqnarray}

Although not very illuminating, this formula may readily be used to recover the 
generating functions $M(P,J)_N$ appropriate to any fixed $N$ by expanding~(\ref{masterGF})
in powers of ${\cal N}$. In particular we find:

\begin{eqnarray}
M(P,J)_{N=1}&=&\frac {1} {(1-JP)(1-P^2)}\\
M(P,J)_{N=2}&=&\frac {1+JP^2} {(1-JP)^2(1- P^2)^3}\\
M(P,J)_{N=3}&=&\frac { 1+3JP^2+P^3-J^2P^3-3JP^4-J^2P^6} {(1-JP)^3(1- P^2)^6}\\
M(P,J)_{N=4}&=&\frac { 
\begin{array}{c} 
(J^{3}P^{10} + J^{3}P^{8} + 6J^{2}P^{8} + 4J^{3}P^{7} - 4JP^{7} 
+ J^{3}P^{6} - 10J^{2}P^{6} + P^{6}\cr
 - 10JP^{4} + P^{4} + J^{3}P^{4} - 4J^{2}P^{3} + 4P^{3} + 6JP^{2} + P^{2} + 1)\cr
\end{array} }
{(1-JP)^4(1- P^2)^9 }
\label{MPJn} 
\end{eqnarray}

Finally, it follows from (\ref{step1}) that the required generating function for $\gl(1|3N)$ 
is obtained by including an additional factor of $(1+P)$ in the numerator
of each of these expressions. For example in the case $N=1$, the $\gl(1|3)$ generating 
function is
\begin{equation}
G(P,J)_{N=1}=(1+P)M(P,J)_{N=1}=\frac{1}{(1-JP)(1-P)}\,,
\label{G13}
\end{equation}
whose expansion takes the form
\begin{equation}
G(P,J)_{N=1}=1 + (1+J)P + (1+J+J^2)P^2 + (1+J+J^2+J^2+J^3)P^3 + \cdots \,.
\label{G13exp}
\end{equation}
This corresponds of course to our earlier observation (\ref{VpN1}) that in the
$N=1$ case $V(p)$ contains states of angular momentum $j=0,1,\ldots,p$,
without multiplicity.

By way of a more interesting example, in the case $N=2$ the complete
generating function is (see also~\cite{RRS})
\begin{equation}
G(P,J)_{N=2}=\frac{(1+P)(1+P^2J)}{(1-P^2)^3(1-PJ)^2}.
\end{equation}
In the expansion of $G(P,J)_{N=2}$, the coefficient of $P^p$ yields the 
angular momentum content of the
$\gl(1|6)$ representation $V(p)$,
\begin{eqnarray}
G(P,J)_{N=2}&=&1 + (1+2J)P+(3+3J+3J^2)P^2 + (3+7J+5J^2+4J^3)P^3\nn\\
&&+ (6+9J+11J^2+7J^3+5J^4)P^4 \nn\\
&&+ (6+15J+15J^2+15J^3+9J^4+6J^5)P^5+\cdots
\end{eqnarray}
For example, for $V(3)$ the possible total angular momentum values are $j=0,1,2,3$ with multiplicities 
$3,7,5$ and $4$ respectively.
Note that in general for $\gl(1|6)$ the total angular momentum values in $V(p)$ are given,
as expected, by $j=0,1,\ldots,p$,
but each value appears with a certain multiplicity.

To conclude, the complete  angular momentum content (including multiplicities) for all representations
$V(p)$ of $\gl(1|3N)$ is resolved by $G(P,J)_N=(1+P) M(P,J)_N$, where $M(P,J)_N$ is the
coefficient of ${\cal N}^N$ in the expansion of the master generating function~(\ref{masterGF}).

\setcounter{equation}{0}
\section{Position operator properties and a classical limit for $V(p)$}

The special class of representations $V(p)$ not only allows a complete analysis of the angular momentum content,
it also allows a deeper analysis of the position and momentum operator properties.
Let us consider in more detail the single particle ($N=1$) $D$-dimensional WQO 
in the Hilbert space $V(p)$ as representation space.
A set of basis vectors is given by
\begin{equation}
v(\theta;r)\equiv v(\theta;r_1,r_2, \ldots,r_D),\qquad
\theta\in\{0,1\},\ r_i\in\{0,1,2,\ldots\},\ \hbox{ and } \theta+r_1+\cdots+r_D= p.
\label{newbasis2}
\end{equation}
There are only two distinct energy levels:
\begin{equation}
\hat{H} v(\theta;r) = \hbar\omega (\frac{p}{D-1}+\theta) v(\theta;r).
\end{equation}
Furthermore, the squares of the position operator components satisfy:
\begin{equation}
\hat R^2_i v(\theta;r) = \frac{\hbar}{(D-1)m\omega} (r_i+\theta) v(\theta;r).
\end{equation}
Similarly, one has
\begin{equation}
\hat P^2_i v(\theta;r) = \frac{\hbar m\omega}{D-1} (r_i+\theta) v(\theta;r).
\end{equation}

In order to illustrate some of the spatial properties, let us first assume that $D=2$, i.e.\
we are dealing with a 2-dimensional one particle WQO.
The squares of the position operator components commute, $[\hat R_1^2,\hat R_2^2]=0$, and the $2p+1$ basis vectors
$v(\theta;r_1,r_2)$ are common eigenvectors:
\begin{eqnarray}
&&\hat R^2_1 v(\theta;r) = \frac{\hbar}{m\omega} (r_1+\theta) v(\theta;r), \quad
\hat R^2_2 v(\theta;r) = \frac{\hbar}{m\omega} (r_2+\theta) v(\theta;r), \nn\\
&&(\hat R^2_1+\hat R^2_2) v(\theta;r) = \frac{\hbar}{m\omega} (p+\theta) v(\theta;r).
\end{eqnarray}
Let us consider in Figure~1 a plot of the eigenvalues of 
$\hat R_1^2$ and $\hat R_2^2$ with respect to these eigenvectors, for some values of $p$.
\begin{figure}[htb]
\caption{Eigenvalues of $\hat R_1^2$ (on the $x$-axis) and $\hat R_2^2$ (on the $y$-axis), 
for $p=5$ and $p=50$ (in units of $\frac{\hbar}{m\omega}$). 
For each commom eigenvector $v(\theta;r_1,r_2)$
the corresponding eigenvalues are plotted.}
\begin{center}
\includegraphics{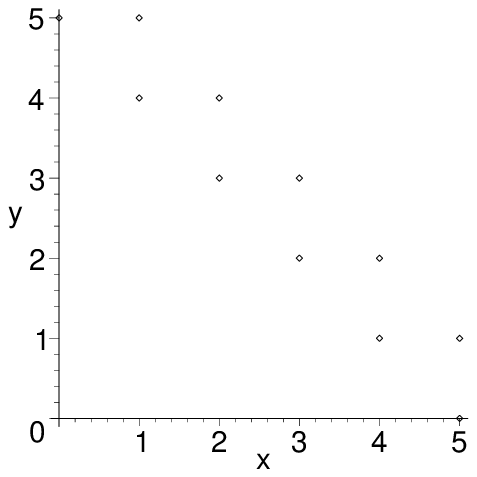}\hskip 15mm
\includegraphics{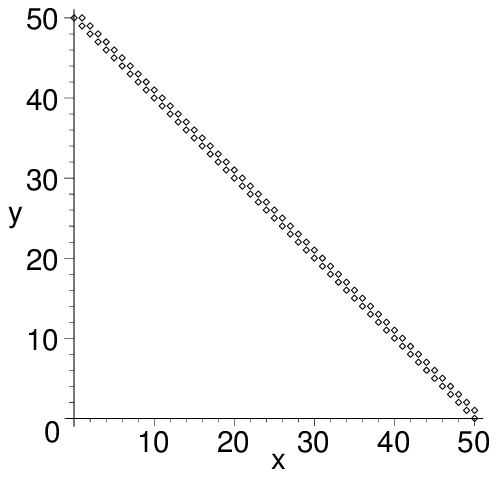}
\end{center}
\end{figure}

Since $\hat R_1$ and $\hat R_2$ do not commute, the position of the oscillating particle cannot be determined. However,
if the system is in one of the basis states $v(\theta;r_1,r_2)$, the above implies that the 
particle can be found in four possible positions, with coordinates
$\sqrt{\frac{\hbar}{m\omega}}(\pm \sqrt{r_1+\theta},\pm \sqrt{r_2+\theta})$ (independent $\pm$-signs).
If we continue in this interpretation, where the positions are determined through measuring the eigenvalues of
$\hat R_1^2$ and $\hat R_2^2$ in their common eigenstates, 
then the possible positions of the particle in the Hilbert space $V(p)$ consists of a collection
of such coordinates $\sqrt{\frac{\hbar}{m\omega}}(\pm \sqrt{r_1+\theta},\pm \sqrt{r_2+\theta})$. 
This leads to Figure~2.
\begin{figure}[htb]
\caption{Possible positions of the WQO, for $p=5$ and $p=50$ (in units of $\sqrt{\frac{\hbar}{m\omega}}$).}
\begin{center}
\includegraphics{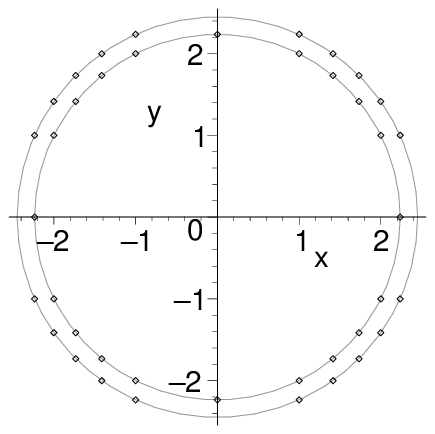}\hskip 15mm
\includegraphics{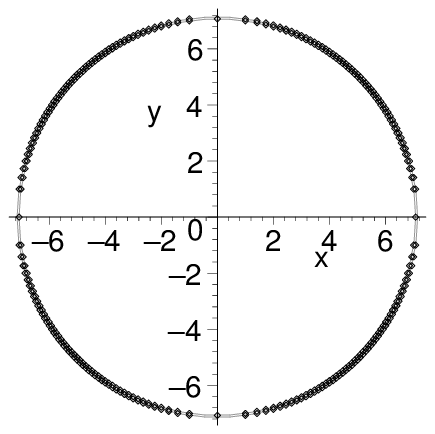}
\end{center}
\end{figure}
Thus the particle is situated on one of two circles, one with radius $\rho_0=\sqrt{\frac{p\hbar}{m\omega}}$ and 
one with radius $\rho_1=\sqrt{\frac{(p+1)\hbar}{m\omega}}$. The spectrum of the coordinate operators is
discrete, however. 

The current situation is interesting, since the classical limit of the WQO in this Hilbert space $V(p)$
can be considered. For this purpose, one should examine the situation where 
the quantum numbers of the system become very large (in the limit to infinity) and $\hbar$ becomes
very small (in the limit to~0). So let us deal with the situation where 
\begin{equation}
p\rightarrow +\infty, \qquad \hbar \rightarrow 0, \qquad\hbox{but}\qquad p\hbar \rightarrow C,
\label{lim}
\end{equation}
for some constant $C$.
We shall examine what happens to the spectrum of the physical operators $\hat H$, $\hat R_1$, $\hat R_2$ and 
$\hat R_1^2+\hat R_2^2$ under this limit. Using~(\ref{lim}),
the energy of the system becomes constant, $C\omega$, as the two energy levels
coincide under the limiting process. 
The limit of the spectrum of $\hat R_1$ and $\hat R_2$ is the interval
$[-\sqrt{\frac{C}{m\omega}},\sqrt{\frac{C}{m\omega}}]$.
The eigenvalue of $\hat R_1^2+\hat R_2^2$ becomes $\frac{C}{m\omega}$, a fixed constant.
So the particle is situated on a circle with radius $\sqrt{\frac{C}{m\omega}}$. 
In the limit, all possible positions on the circle are possible. 
This suggests that under this limit the solution of the WQO corresponds to one of the classical solutions
\begin{equation}
x(t)=A\cos(\omega t), \qquad y(t)=A\sin(\omega t)
\end{equation}
of a 2-dimensional simple homogeneous harmonic oscillator, determined by the differential equations
\begin{equation}
\ddot{x}(t) + \omega^2\, x(t) =0, \qquad \ddot{y}(t) + \omega^2\, y(t) =0,
\end{equation}
where $\omega^2=\frac{k}{m}$ ($k$ being the ``spring constant''). 
The classical energy of such a system is $m\omega^2A^2$, and the classical
oscillator moves on a circle with radius $A$. Under the identification of the constants $C=m\omega A^2$, 
both the energy and position of the WQO (for this solution) are in agreement with those of 
the particular solution of the classical oscillator. In other words, the ``correspondence principle''
holds for this solution of the WQO.

It should be clear that the above considerations about possible particle positions
remain valid for arbitrary $D$. For instance, when $D=3$, the possible positions
of the particle in one of the basis states of $V(p)$ consists of a collection of coordinates 
$\sqrt{\frac{\hbar}{m\omega}}(\pm \sqrt{r_1+\theta},\pm \sqrt{r_2+\theta},\pm \sqrt{r_3+\theta})$ 
(independent $\pm$-signs). The collection of such coordinates leads to Figure~3,
i.e.\ a set of points on two concentric spheres, one with radius $\sqrt{\frac{p\hbar}{2m\omega}}$
and one with radius $\sqrt{\frac{(p+1)\hbar}{2m\omega}}$.
\begin{figure}[htb]
\caption{Possible positions of the 3-dimensional WQO, for $p=5$ (in units of $\sqrt{\frac{\hbar}{2m\omega}}$).
In the left figure, the set of points is on a sphere with radius $\sqrt{p+1}$; 
in the right figure, they are on a sphere with radius $\sqrt{p}$.}
\begin{center}
\includegraphics{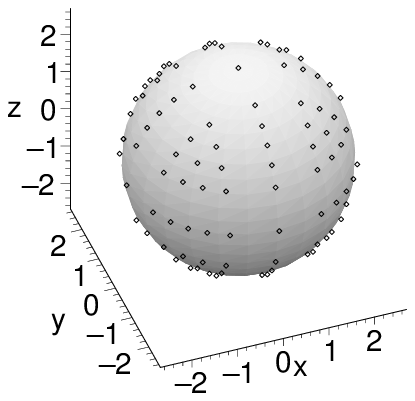}\hskip 15mm
\includegraphics{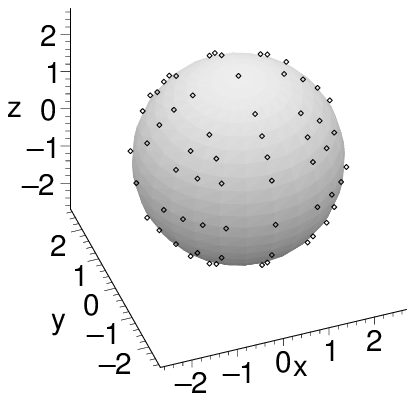}
\end{center}
\end{figure}

As far as the position or momentum operators for the WQO is concerned, we have not yet considered their
time dependence. Following~\cite{Palev5}-\cite{K2}, this is given by
\begin{equation}
\hat R_k(t)=\sqrt{\frac{\hbar}{(D-1)m\omega}}(A^+_k e^{-i\omega t} + A^-_k e^{i\omega t}), \quad
\hat P_k(t)=-i\sqrt{\frac{m\omega\hbar}{(D-1)}}(A^+_k e^{-i\omega t} - A^-_k e^{i\omega t}), 
\end{equation}
for the $D$-dimensional single particle case.
For a general mixed state of $V(p)$,
\begin{equation}
|x\rangle = \sum_{\theta,r} c_{\theta;r} v(\theta;r),\qquad
\sum_{\theta,r} c_{\theta;r}^2=1,
\end{equation}
one finds that the average value or mean trajectories of the coordinate and momentum operator components is
\begin{equation}
\langle x | \hat R_k | x\rangle =
2 \sqrt{\frac{\hbar}{(D-1)m\omega}} B_k \cos(\omega t), \qquad
\langle x | \hat P_k | x\rangle =
-2 \sqrt{\frac{m\omega\hbar}{D-1}} B_k \sin(\omega t),
\end{equation}
where $B_k$ is the constant
\begin{equation}
B_k= \sum_r \sqrt{r_k} c_{0;r_1,\ldots,r_D} c_{1;r_1,\ldots,r_k-1,\ldots,r_D}.
\end{equation}
If the system is in a stationary state (fixed energy eigenvalue), one has either all $c_{0,r}=0$ or else
all $c_{1;r}=0$. In that case, all $B_k$ are zero. In other words, for stationary states $|x\rangle$ the
average values $\langle x | \hat R_k | x\rangle$ and $\langle x | \hat P_k | x\rangle$ vanish.

The analysis given so far was for a single particle $D$-dimensional WQO in the representation space $V(p)$
of $\gl(1|D)$. One can also consider the same representations 
$V(p)$ of $gl(1|N)$ 
as state spaces of a system of $N$ 1-dimensional
WQO's ($N>1$, since a single 1-dimensional WQO does not have a solution in terms of $\gl(1|1)$).
In this second interpretation, the basis states
\begin{equation}
v(\theta;r)\equiv v(\theta;r_1,r_2, \ldots,r_N),\qquad
\theta\in\{0,1\},\ r_i\in\{0,1,2,\ldots\},\ \hbox{ and } \theta+r_1+\cdots+r_N= p.
\label{newbasis3}
\end{equation}
are the same as in~(\ref{newbasis2}), and similarly
\begin{equation}
\hat{H} v(\theta;r) = \hbar\omega (\frac{p}{N-1}+\theta) v(\theta;r), \qquad
\hat R^2_i v(\theta;r) = \frac{\hbar}{(N-1)m\omega} (r_i+\theta) v(\theta;r).
\end{equation}
In other words, all the previously obtained formulas remain valid, only the interpretation of the
system is different: now it consists of $N$ one-dimensional oscillating particles, and $\hat R_i$ is
the operator corresponding to the position of particle~$i$.
In the representation space $V(p)$, the spectrum of the operator $\hat R_i^2$ is $\{0,1,\ldots,p\}$
and that of $\hat R_i$ is $\{0,\pm{1},\pm\sqrt{2},\ldots, \pm\sqrt{p}\}$. 
Let us again assume that $N=2$, so that we are dealing with
just two one-dimensional oscillating
particles. If the system is in one of the basis states $v(\theta;r_1,r_2)$, our analysis
leads to the interpretation that the position of the first one-dimensional oscillator is 
$\pm \sqrt{\frac{\hbar(r_1+\theta)}{m\omega}}$, and that of the second one-dimensional oscillator 
is $\pm \sqrt{\frac{\hbar(r_2+\theta)}{m\omega}}$. Since $\theta+r_1+r_2=p$, these positions
are not independent but ``correlated''. 
To see this from a different point of view, 
recall that the positions of the two oscillators cannot be measured simultaneously, since 
$[\hat R_1,\hat R_2]\ne 0$. But suppose the system is in an arbitrary state of $V(p)$, and the position
of the first oscillator is measured at time~$t=0$. 
This measurement has one of the eigenvalues $\pm \sqrt{\frac{\hbar r_1}{m\omega}}$ of $\hat R_1(t)$
as an outcome, with $r_1\in\{0,1,\ldots,p\}$.
As a consequence of the measurement, the system is then in an eigenstate of $\hat R_1$ for this eigenvalue.
It is easy to verify that this eigenstate is unique, given by (we assume $r_1>0$)
\begin{equation}
\frac{1}{\sqrt{2}} v(0;r_1, r_2) \pm \frac{1}{\sqrt{2}} v(1; r_1-1, r_2),
\end{equation}
with $r_2=p-r_1$ (the plus sign for the positive eigenvalue,
and the minus sign for the negative eigenvalue). 
The time evolution of this system is then described by
\begin{equation}
|x\rangle = \frac{e^{i\omega t/2}}{\sqrt{2}} v(0;r_1, r_2) \pm \frac{e^{-i\omega t/2}}{\sqrt{2}} v(1; r_1-1, r_2).
\label{state_x}
\end{equation}
Note that $|x\rangle$ is an eigenstate of $\hat R_1(t)$ for the eigenvalue 
$\pm \sqrt{\frac{\hbar r_1}{m\omega}}$.
The state~(\ref{state_x}) can be rewritten as follows:
\begin{eqnarray}
|x\rangle &=& \frac{1}{2\sqrt{2}} [e^{i\omega t/2}v(0;r_1,r_2)+e^{-i\omega t/2}v(1;r_1,r_2-1)]\nn\\ 
&&+ \frac{1}{2\sqrt{2}} [e^{i\omega t/2}v(0;r_1,r_2)- e^{-i\omega t/2}v(1;r_1,r_2-1)] \nn\\
&& \pm \frac{1}{2\sqrt{2}} [e^{i\omega t/2}v(0;r_1-1,r_2+1)+e^{-i\omega t/2}v(1;r_1-1,r_2)]\nn\\
&&\mp \frac{1}{2\sqrt{2}} [e^{i\omega t/2}v(0;r_1-1,r_2+1)- e^{-i\omega t/2}v(1;r_1-1,r_2)].
\end{eqnarray}
The four linear combinations in square brackets are four eigenvectors
of $\hat R_2(t)$, with eigenvalues $\sqrt{\frac{\hbar}{m\omega}}\sqrt{p-r_1}$, 
$-\sqrt{\frac{\hbar}{m\omega}}\sqrt{p-r_1}$, $\sqrt{\frac{\hbar}{m\omega}}\sqrt{p-r_1+1}$ and 
$-\sqrt{\frac{\hbar}{m\omega}}\sqrt{p-r_1+1}$ respectively.
This implies that a measurement of the position of the first oscillator, i.e.\ of 
$\hat R_1$ at time~$t=0$ with outcome $\sqrt{\frac{\hbar r_1}{m\omega}}$, imposes strong
restrictions on the position eigenvalues of the second oscillator, measured at time $t$. 
The two one-dimensional oscillators are not free but correlated in some sense.

Let us also examine the classical limit~(\ref{lim}) for this second interpretation.
In fact, the limit of the spectrum of the operators $\hat H$, $\hat R_1$, $\hat R_2$ and 
$\hat R_1^2+\hat R_2^2$ is the same as before. So the total
energy of the system becomes constant, $C\omega$.
The limit of the spectrum of $\hat R_1$ and of $\hat R_2$ is the interval
$[-\sqrt{\frac{C}{m\omega}},\sqrt{\frac{C}{m\omega}}]$, and
the eigenvalue of $\hat R_1^2+\hat R_2^2$ is fixed, $\frac{C}{m\omega}$.
Thus, under this limit the solution of the 2-particle 1-dimensional WQO corresponds to
the classical solution
\begin{equation}
x_1(t)=A\cos(\omega t), \qquad x_2(t)=A\sin(\omega t).
\end{equation}
This simply describes a system of two uncoupled identical 1-dimensional harmonic oscillators,
but with a fixed phase difference in their oscillations. In other words, the initial conditions
of the two oscillators are related: also here there is a correlation.
So the ``correspondence principle'' also holds for this interpretation of the WQO.

\setcounter{equation}{0}
\section{Analysis for a general example}

The results of the previous sections were mainly devoted to the special representations $V(p)$.
However, the computations of Section~2 allow us to make an analysis of the WQO properties for
any unitary representation. 
As an example, we consider a representation $W([m]_{3N+1})$ which
belongs to neither the set $W(p)$ nor the set $V(p)$,
namely the representation corresponding to the $\gl(1|3N)$ 
irreducible representation $V^{\{2,1\}}_{\gl(1|3N)}$ with highest weight 
$\Lambda=2\epsilon+\delta_1$. This can be identified 
with the $\gl(1|3N)$ module $W([m]_{3N+1})$ with $[m]_{3N+1}=[2,1,0,\ldots,0]$.

The branching from $\gl(1|3N)$ to $\gl(1)\otimes \gl(3N)$ is given by
\begin{equation}
     V^{\{2,1\}}_{\gl(1|3N)}\rightarrow 
     V^{2}_{\gl(1)}\times V^{\{1\}}_{\gl(3N)} 
     +V^{1}_{\gl(1)}\times V^{\{2\}}_{\gl(3N)} 
     +V^{1}_{\gl(1)}\times V^{\{1,1\}}_{\gl(3N)} 
     +V^{0}_{\gl(1)}\times V^{\{2,1\}}_{\gl(3N)} \,.
\label{V21}
\end{equation} 
The four terms arising here are associated with the 
four sets of states:
\begin{equation}
{\scriptsize
\left|
\begin{array}{lcllll}
2 & 1 & 0 & 0 & \cdots & 0  \\
  & 1 & 0 & 0 & \cdots & 0  \\
  & * & * & 0 & \cdots &  \\
  &\vdots & \vdots &\qdots & & \\
& * & * & & & \\
& * & & & &
\end{array}
\right)
\quad
\left|
\begin{array}{lcllll}
2 & 1 & 0 & 0 & \cdots & 0  \\
  & 1 & 1 & 0 & \cdots & 0  \\
  & * & * & 0 & \cdots &    \\
  &\vdots & \vdots & \qdots & & \\
& * & * & & & \\
& * & & & &
\end{array}
\right)
\quad
\left|
\begin{array}{lcllll}
2 & 1 & 0 & 0 & \cdots & 0  \\
  & 2 & 0 & 0 & \cdots & 0  \\
  & * & * & 0 & \cdots &   \\
  &\vdots & \vdots & \qdots  & & \\
& * & * & & & \\
& * & & & &
\end{array}
\right)
\quad
\left|
\begin{array}{lcllll}
2 & 1 & 0 & 0 & \cdots & 0  \\
  & 2 & 1 & 0 & \cdots & 0  \\
  & * & * & 0 & \cdots &    \\
  &\vdots & \vdots & \qdots & & \\
& * & * & & &\\
& * &   & & &
\end{array}
\right)
}
\label{21m}
\end{equation}
with the entries $*$ taking values determined by the various 
betweennness conditions. 
For all states within each of these four sets the corresponding 
values of $(\theta_1,\theta_2,\ldots)$ are given by
$(0,0,0,\dots,0)$, $(0,1,0,\ldots,0)$, $(1,0,0,\ldots,0)$ and $(1,1,0,\dots,0)$,
respectively. In the notation of (\ref{Eq}) we have
\begin{equation}
   E_q=\hbar \omega \left( \frac{6N+1}{3N-1} -q \right)
\label{Eq21}
\end{equation}
with $q=0$, $1$, $1$ and $2$, for our four sets of states. It follows that
we have just three distinct energy levels with degeneracies as shown 
below:
\begin{equation}
\begin{array}{ll}
   E_0=\hbar\omega \ds\frac{6N+1}{3N-1}\,,&\quad d_0=3N\,;\cr\cr
   E_1=\hbar\omega \ds\frac{3N+2}{3N-1}\,,&\quad d_1=9N^2\,;\cr\cr
   E_2=\hbar\omega \ds\frac{3}{3N-1}\,,   &\quad d_2=N(9N^2-1)\,.\cr
\end{array}   
\label{E012}
\end{equation}
These correspond precisely to the three $\gl(1)$ representations, $V^{2}_{\gl(1)}$, 
$V^{1}_{\gl(1)}$ and $V^{0}_{\gl(1)}$ appearing in (\ref{V21}). The degeneracies,
$d_q$ have been obtained by using $N$-dependent formulae for the dimensions of 
the irreducible representations of $\gl(3N)$ that appear in (\ref{V21}). 

The spectrum of $\hR^2_{\a i}$ may be obtained from (\ref{actionR2}) by 
setting $n=3N$ and applying the formula to the above states (\ref{21m}).
One finds that for states of these four types the factor 
$(m_{0,n+1}+\ldots+m_{n,n+1})-(m_{1,n}+\cdots+m_{n,n})$ takes
the values $2$, $1$, $1$ and $0$, respectively, while
the second factor $(m_{1,k}+\cdots+m_{k,k})-(m_{1,k-1}+\cdots+m_{k-1,k-1})$
lies in the sets $\{0,1\}$, $\{0,1\}$, $\{0,1,2\}$ and $\{0,1,2\}$,
respectively. It follows that for each of the possible energy 
levels $E_q$ the eigenvalues of $\hR^2_{\a i}$ are given by
\begin{equation}
   R^2_{\a i}=\frac{\hbar\ r^2_{\a i}}{(3\,N-1)\,m\,\omega}
   \hbox{~~with~~} r^2_{\a i} = \bigg\{
   \begin{array}{ll}
       2,3&\hbox{~for~$q=0$}\,;\cr
       1,2,3&\hbox{~for~$q=1$}\,;\cr
       0,1,2&\hbox{~for~$q=2$}\,.\cr
   \end{array}
\label{R012}
\end{equation}
Note that, unlike all the $V(p)$ cases with $p>0$, there exists a configuration with
$r_{\a i}^2=0$, namely one of the $q=2$ ground state configurations.
 
To determine the angular momentum content of these states, we can
extend (\ref{V21}) by considering the chain of subalgebras:
\begin{equation}
    \gl(3N)\rightarrow \gl(3) \oplus \gl(N)
\rightarrow \so(3) \oplus \gl(N)
\rightarrow \so(3).
     \label{gl3n}
\end{equation} 
The corresponding branchings take the form:

\begin{eqnarray}
\hskip-1cm
&V^{\{1\}}_{\gl(3N)}
&\rightarrow V^{\{1\}}_{\gl(3)} \times V^{\{1\}}_{\gl(N)}\cr
\hskip-1cm
&&\rightarrow V^{1}_{\so(3)} \times V^{\{1\}}_{\gl(N)}\cr
\hskip-1cm
&&\rightarrow N\, V^{1}_{\so(3)}\\ \cr 
\hskip-1cm
&V^{\{2\}}_{\gl(3N)}
&\rightarrow V^{\{2\}}_{\gl(3)} \times V^{\{2\}}_{\gl(N)}+
V^{\{1,1\}}_{\gl(3)} \times V^{\{1,1\}}_{\gl(N)}\cr
\hskip-1cm
&&\rightarrow (V^{2}_{\so(3)}+V^{0}_{\so(3)})\times V^{\{2\}}_{\gl(N)}  
  +V^{1}_{\so(3)}\times V^{\{1,1\}}_{\gl(N)}\cr
\hskip-1cm
&&\rightarrow  \frac12\, N(N+1)\,V^{2}_{\so(3)}
+ \frac12 N(N-1)\,V^{1}_{\so(3)}
+\frac12 N(N+1)\,V^{0}_{\so(3)} \\ \cr
\hskip-1cm
&V^{\{1,1\}}_{\gl(3N)}
&\rightarrow V^{\{2\}}_{\gl(3)} \times V^{\{1,1\}}_{\gl(N)}+
V^{\{1,1\}}_{\gl(3)} \times V^{\{2\}}_{\gl(N)}\cr
\hskip-1cm
&&\rightarrow (V^{2}_{\so(3)}+V^{0}_{\so(3)})\times V^{\{1,1\}}_{\gl(N)}  
  +V^{1}_{\so(3)}\times V^{\{2\}}_{\gl(N)}\cr 
\hskip-1cm
&&\rightarrow  \frac12 N(N-1)\,V^{2}_{\so(3)}
+ \frac12 N(N+1)\,V^{1}_{\so(3)}+\frac12 N(N-1)\,V^{0}_{\so(3)} \\ \cr
\hskip-1cm
&V^{\{2,1\}}_{\gl(3N)}
&\rightarrow V^{\{3\}}_{\gl(3)} \times V^{\{2,1\}}_{\gl(N)}
+V^{\{2,1\}}_{\gl(3)}\times (V^{\{3\}}_{\gl(N)}+V^{\{2,1\}}_{\gl(N)}+ V^{\{1,1,1\}}_{\gl(N)})
+V^{\{1,1,1\}}_{\gl(3)}\times V^{\{2,1\}}_{\gl(N)}\cr
\hskip-1cm
&&\rightarrow (V^{3}_{\so(3)}+ V^{1}_{\so(3)})\times V^{\{2,1\}}_{\gl(N)}+
(V^{2}_{\so(3)}+ V^{1}_{\so(3)})\times (V^{\{3\}}_{\gl(N)}+V^{\{2,1\}}_{\gl(N)}+ V^{\{1,1,1\}}_{\gl(N)})\cr 
\hskip-1cm
&&\qquad + V^{0}_{\so(3)}\times V^{\{2,1\}}_{\gl(N)}\cr
\hskip-1cm
&&\rightarrow \frac13 N(N^2-1)\,V^{3}_{\so(3)}+\frac13 N(2N^2+1)\,V^{2}_{\so(3)}
+ N^3\,V^{1}_{\so(3)}+ \frac13 N(N^2-1)\,V^{0}_{\so(3)} \,.
\label{br3N}  
\end{eqnarray}
Combining these results with (\ref{V21}) we can write down the generating function
\begin{eqnarray}
G(Q,J)_N &=& \sum_{q,j=0}^\infty n_{q,j}(N)\, Q^q\,J^j \cr
& =& N\,J\,Q^0 + N^2\,(1+J+J^2)\,Q^1  \cr 
&& + \left(\frac13 N(N^2-1)+ N^3\,J+\frac13 N(2N^2+1)\,J^2
+ \frac13 N(N^2-1)\,J^3 \right)\,Q^2
\label{GQJ}
\end{eqnarray}
where the coefficient $n_{q,j}(N)$ gives the multiplicity 
of states of angular momentum $j$ and energy $E_q$ as a function of $N$.  
To recover the degeneracy of each energy level one replaces each $J^j$ by 
the corresponding dimension $2j+1$ of $V^j_{\so(3)}$ to give
\begin{equation}
G(Q)_N= 3N\,Q^0+9N^2\,Q^1+ N(9N^2-1)\,Q^2,
\label{GQ}
\end{equation} 
in agreement with (\ref{E012}).
Similarly, the full angular momentum content for specific values of $N$ may
be obtained from~(\ref{GQJ}) by setting $Q=1$. For example, we obtain
\begin{eqnarray}
G(1,J)_{N=1} &=& 1+3J+2J^2; \\
G(2,J)_{N=2} &=& 6+14J+10J^2+2J^3.
\end{eqnarray}

Although this example is not very remarkable, it illustrates how the methods 
constructed here can be used, and it shows that the material developed in this
paper can in principle be applied to any unitary representation or class of unitary representations.

\setcounter{equation}{0}
\section{Conclusions}

The theory of WQOs has a deep connection with representation theory of Lie superalgebras.
Solutions of WQO systems can be considered not only for the Lie superalgebra $\ssl(1|n)$,
but also for other Lie superalgebras. As typical examples, we mention the $\osp(3|2)$ 
WQO~\cite{Palev6}
and the recently studied $\ssl(3|N)$ WQO~\cite{Palev7}. 
The physical properties of the WQO depend on the Lie superalgebra, and on the class of
representations considered.

This paper has made a contribution to the study of physical properties of WQOs of type $\ssl(1|n)$,
initiated in~\cite{Palev1} and performed in detail for the 3-dimensional WQO 
in~\cite{Palev5}-\cite{K2}.
First of all, our current approach is slightly more general by considering the $D$-dimensional
$N$-particle WQO of type $\ssl(1|n)$.
Secondly, in earlier papers only the solutions corresponding to Fock representations $W(p)$
were investigated.
Here, we have initiated the study of solutions related to all (unitary) representations of $\ssl(1|DN)$.
This requires the construction of a proper basis (the GZ-basis) for these representations, with
explicit actions of the $\ssl(1|DN)$ generators, as given here in Section~2.

In order to emphasize the difference with the previously  considered Fock spaces $W(p)$, we have 
paid  particular attention to another special class of representations $V(p)$. 
These representations are, once again, weight multiplicity free, 
so they are easy to describe explicitly. The energy spectrum and the spectra of
position and momentum operators in these representations have been  obtained without much effort. 
Among the special features we mention: there are only two different energy levels (the difference 
being $\hbar\omega$); the position operator components are non-commuting operators and have a discrete
spectrum, 
leading to
spatial properties that are quite different from those of a canonical
quantum oscillator.

In addition, for these representations $V(p)$ the complete angular momentum content has
been obtained in the form of a generating function using powerful group theoretical methods.

Another particularly nice result is that, for those solutions related to
$V(p)$, a classical limit has been obtained,
 showing that the behavior of the WQO solution reduces to a classical solution
of the harmonic oscillator. 
For the previously considered Fock space
solutions based on $W(p)$, such a limit does not exist.

Wigner Quantum Oscillators, belonging to the class of non-canonical quantum systems, 
possess many attractive and fascinating properties, both for mathematicians and physicists. 
We believe that the correspondence
principle observed here makes them even more interesting for physics.

\section*{Acknowledgments}
The authors would like to thank Professor T.D. Palev for his interest.
NIS was supported by a project from the Fund for Scientific Research -- Flanders (Belgium).
In addition, RCK is grateful for the
award of a Leverhulme Emeritus Fellowship in support of this work.


\begin{thebibliography}{99}

\bibitem{Kac1}
Kac V G 1977 {\it Adv.\ Math.} {\bf 26} 8
\bibitem{Kac2}
Kac V G 1978 {\it Lect.\ Notes Math.} {\bf 676} 597 
\bibitem{Palev2}
Palev T D 1989 {\it J.\ Math.\ Phys.} {\bf 30} 1433
\bibitem{Wigner}
Wigner E P 1950 {\it Phys. Rev.} {\bf 77} 711 
\bibitem{Palev1}
Palev T D 1982 {\it J.\ Math.\ Phys.} {\bf 23} 1778; 1982 {\it Czech.\ J.\ Phys.} 
{\bf B32} 680
\bibitem{Palev4}
Kamupingene A H, Palev T D  and Tsaneva S P 1986 {\it J.\ Math.\ Phys.} {\bf 27} 2067
\bibitem{Palev5}
Palev T D and Stoilova N I 1997 {\it J.\ Math.\ Phys.}
        {\bf 38} 2506 
\bibitem{K1}
King R C, Palev T D, Stoilova N I and Van der Jeugt J 2003 
{\it J.\ Phys.\ A: Math.\ Gen.} {\bf 36} 4337 
\bibitem{K2}
King R C, Palev T D, Stoilova N I and Van der Jeug J 2003 
{\it J.\ Phys.\ A: Math.\ Gen.} {\bf 36} 11999
\bibitem{SJ}
Stoilova N I and Van der Jeugt J 2005
{\it J.\ Phys.\ A: Math.\ Gen.} {\bf 38} 9681
\bibitem{Gould} 
Gould M D and Zhang R B  1990 {\it J.\ Math.\ Phys.} {\bf 31} 2552 
\bibitem{Palev3}
Palev T D 1980 {\it J.\ Math.\ Phys.} {\bf 21} 1293 
\bibitem{HS}
Hatzinikitas A and Smyrakis I {\it J.\ Math.\ Phys.} 2002
{\bf 43}, 113 
\bibitem{C}
Connes A 1994 {\it Non-commutative geometry} (San Diego: Academic Press)
\bibitem{NP}
Nair V P and Polychronakos A P 2001 {\it Phys.\ Lett.} {\bf B 505} 267
\bibitem{SS1}
Smailagic  A and Spallucci E 2002 {\it Phys.\ Rev.}
{\bf D 65}: 107701
\bibitem{SS2}
Smailagic A and Spallucci E  2002 {\it J.\ Phys.\ A}
{\bf  35}, L363 
\bibitem{MM}
Mathukumar  B and Mitra P 2002 {\it Phys.\ Rev.}
{\bf D 66}: 027701 
\bibitem{Molien}
Molien T 1897 {\it Uber die Invarianten der linearen Substitutionsgruppe}
(Sitzungsber. Konigl. Preuss. Akad. Wiss.) 1152
\bibitem{Sturmfels}
Sturmfels B 1993 {\it Algorithms in Invariant Theory}
(Wien: Springer-Verlag) p. 188
\bibitem{LJ} 
Bernstein I N and Leites D A 1980 {\it C.R. Acad. Bulg. Sci.} {\bf 33} 1049
\bibitem{JHKR}
Van der Jeugt J, Hughes J W B, King R C and Thierry-Mieg J 1990 {\it 
Commun.\ Alg.} {\bf 18} 3453 
\bibitem{GZ}
Gel'fand I M and Zetlin M L 1950 {\it Dokl.\ Akad.\ Nauk SSSR} {\bf 71} 825
\bibitem{JHKT}
Van der Jeugt J, Hughes J W B, King R C and Thierry-Mieg J 1990
{\it J\ Math\ Phys} {\bf 31} 2278
\bibitem{Schur}
SCHUR, an interactive program for calculating the properties of Lie groups and 
symmetric functions, distributed by S Christensen. 
\bibitem{K3}
King R C 1975 {\it J.\ Phys. A: Math. Gen.} {\bf 8} 429 
\bibitem{RRS}
Raychev P P, Roussev, R P and Smirnov Yu F 1991
{\it J. Phys. A: Math. Gen.} {\bf 24}  2943
\bibitem{Palev6}
Palev T D and Stoilova N I 1994 {\it J.\ Phys. A}
        {\bf 27} 977; 7387
\bibitem{Palev7}
Palev T D 2006 $SL(3|N)$ Wigner quantum oscillator: examples of ferromagnetic-like
oscillators with noncommutative, square-commutative geometry {\it Preprint} hep-th/0601201  


        
\end{thebibliography}
\end{document}